\documentclass[12pt]{iopart}
 \usepackage{cite}
\usepackage{gensymb}
\usepackage{iopams}
\usepackage{amssymb,amsfonts}
\usepackage{graphicx}
\usepackage{verbatim}
\usepackage{tabularx} 
\usepackage{multirow} 

\usepackage{hyperref} 

\def\BibTeX{\text{B\kern-.05em{\sc i\kern-.025em b}\kern-.08em
    T\kern-.1667em\lower.7ex\hbox{E}\kern-.125emX}}

\begin{document}

begin{document}

\title[Heterogeneous Integration of...]{Heterogeneous integration of superconducting thin films and epitaxial semiconductor heterostructures with Lithium Niobate}

\author{Michelle Lienhart$^{1,4,*}$, Michael Choquer$^{2,*}$, Emeline D.S. Nysten$^{3,1,*}$, Matthias Weiß$^{3,1}$, Kai M\"{u}ller$^{5,6}$, Jonathan J. Finley$^{4,6}$, Galan Moody$^2$, Hubert J. Krenner$^{3,1}$}

\def\BibTeX{\text{B\kern-.05em{\sc i\kern-.025em b}\kern-.08em
    T\kern-.1667em\lower.7ex\hbox{E}\kern-.125emX}}


\address{$^1$Lehrstuhl f\"{u}r Experimentalphysik 1 and Augsburg Centre for Innovative Technologies (ACIT), Universit\"{a}t Augsburg, 86150 Augsburg, Germany}
\address{$^2$Electrical and Computer Engineering Department, University of California, Santa Barbara, California 93106, USA}
\address{$^3$Physikalisches Institut, Westf\"{a}lische Wilhelms-Universit\"{a}t M\"{u}nster, 48149 M\"{u}nster, Germany}
\address{$^4$Walter Schottky Institut, School of Natural Sciences, Technische Universität München, 85748 Garching, Germany}
\address{$^5$Walter Schottky Institut, School of Computation, Information and Technology, Technische Universität München, 85748 Garching, Germany}
\address{$^6$Munich Center for Quantum Science and Technology (MCQST), Schellingstr. 4, 80799 Munich, Germany.}
\address{*  contributed equally}
\ead{krenner@uni-muenster.de and moody@ucsb.edu}
\vspace{10pt}


\begin{abstract}

We report on scalable heterointegration of superconducting electrodes and epitaxial semiconductor quantum dots on strong piezoelectric and optically nonlinear lithium niobate. The implemented processes combine the sputter-deposited thin film superconductor niobium nitride and III-V compound semiconductor membranes onto the host substrate. The superconducting thin film is employed as a zero-resistivity electrode material for a surface acoustic wave resonator with internal quality factors $Q \approx 17000$ representing a three-fold enhancement compared to identical devices with normal conducting electrodes. Superconducting operation of $\approx 400\,\mathrm{MHz}$ resonators is achieved to temperatures $T>7\,\mathrm{K}$ and electrical radio frequency powers $P_{\mathrm{rf}}>+9\,\mathrm{dBm}$. Heterogeneously integrated single quantum dots couple to the resonant phononic field of the surface acoustic wave resonator operated in the superconducting regime. Position and frequency selective coupling mediated by deformation potential coupling is validated using time-integrated and time-resolved optical spectroscopy. Furthermore, acoustoelectric charge state control is achieved in a modified device geometry harnessing large piezoelectric fields inside the resonator. The hybrid quantum dot - surface acoustic wave resonator can be scaled to higher operation frequencies and smaller mode volumes for quantum phase modulation and transduction between photons and phonons via the quantum dot. Finally, the employed materials allow for the realization of other types of optoelectronic devices, including superconducting single photon detectors and integrated photonic and phononic circuits.
\end{abstract}

\maketitle
\section{Introduction}
\label{sec:intro}

Lithium niobate (LiNbO$_3$) is a key material in the interconnected fields of nonlinear optics and acoustics due to its strong optical nonlinearity (e.g. $\chi^{(2)},\,r_{33}=3\times10^{-11}\mathrm{\,m\,V^{-1}}$) and piezoelectricity (e.g. $d_{22}=21\,\mathrm{pC\, N^{-1}}$) \cite{Weis1985,Zhu2021,Delsing2019}. LiNbO$_3$'s strong electro- and acousto-optical effects can be harnessed in quantum integrated photonic circuits \cite{desiatov_ultra-low-loss_2019,wan2022}. This unique combination of properties renders LiNbO$_3$ a versatile host to synergistically combine various nonlinear photonic elements. Despite these advantages, LiNbO$_3$ does not host optical emitters, which are required for many photonics applications. To overcome this shortcoming, several approaches are the focus of current research, including doping with rare earth ions \cite{Baumann1996, Yuechen2022}, direct growth \cite{Preciado2015} or transfer of 2D semiconductors \cite{Lazic2019,scolfaro_acoustically_2021,patel2022surface}, as well as heterogeneous integration of epitaxial semiconductors \cite{Nysten2017, aghaeimeibodi2018}.

One such method for integrating LiNbO$_3$ with optically active quantum emitters is with piezoelectrically generated surface acoustic waves (SAWs), which mechanically deform the host material and have an associated electric field in piezoelectric materials. SAWs are highly appealing because of their universal coupling to dissimilar quantum systems \cite{Delsing2019,Schuetz2015}. SAW devices are fabricated in a planar manner using straightforward CMOS device processing techniques, which has enabled optomechanical transduction between solid-state quantum emitters \cite{Gell2008,Weiss2018,Choquer2022}, integrated photonic and phononic circuits \cite{Fuhrmann2011,Tadesse2014,Balram2016, Buehler2022}, and many other demonstrations of SAW coupling to optically or electrically active quantum systems sensitive to strain or electric fields \cite{Gustafsson2014,Schulein2015,Golter2016,Satzinger2018,maity_coherent_2020}. Another advantage of SAW devices is that they typically operate at gigahertz (GHz) frequencies, where they can be cooled to the phononic ground-state without active cooling protocols\cite{Schuetz2015, Choquer2022}.  

For quantum applications, SAW devices have to be further improved by mitigating loss channels. These include Ohmic heating in conducting electrodes of finite electrical conductivity. To alleviate these losses, SAW devices can be fabricated using superconducting electrode materials. For example, aluminum (Al) electrodes have been employed in SAW resonators of high quality factors $Q \approx 10^5$ \cite{magnusson_surface_2015, manenti_circuit_2017, andersson_squeezing_2022} proving the feasibility of this strategy. However, superconducting operation of such devices is limited to temperatures below aluminum's critical temperature $T_{\mathrm{c}} \approx 1.2\,\mathrm{K}$.
Such low temperatures are not strictly necessary for quantum control schemes of optically active solid-state two-level systems like semiconductor quantum dots (QDs) or defect centers. These QDs exhibit low decoherence already at moderate temperatures $T \geq 4\,K$, which are accessible with conventional $^4\mathrm{He}$ cryostats. This low decoherence was impressively demonstrated by the implementation of all-optical coherent control schemes \cite{Fras2016, Becker2016}. These experiments are conducted under significantly relaxed conditions, i.e., higher operation temperatures compared to superconducting Al quantum systems. Thus, materials with moderately high $T_{\mathrm{c}}$ would mark a significant advantage for hybrid SAW-QD devices. An additional challenge for integrating single quantum emitters with hybrid phononic quantum technologies remains the enhancement of the interaction strength between QDs and SAW phonons \cite{Metcalfe2010, Weiss2021, Wigger2021a, imany2022quantum} and at the same time mitigating the aforementioned losses due to Ohmic heating by employing superconducting electrodes. The first has motivated the development of hybrid systems consisting of strong piezoelectric SAW substrates with heterogeneously integrated III-V compound semiconductors \cite{Rotter1997, rotter1999, govorov2000}. Heterogeneously integrated SAW devices are an extremely active field of current research, and recently large-scale radio frequency acoustoelectric devices have been realized \cite{Hackett2021, Hackett2023} This versatile approach is naturally suited to realize hybrid QD-SAW devices  with transferred epitaxially grown QD layers \cite{Pustiowski2015,Nysten2017,Nysten2020}.
For the latter, nitride-based superconductors are a leading material platform in the moderate temperature range because these materials exhibit much higher critical temperatures compared to Al. For example, niobium nitride (NbN) exhibits its superconducting transition at  $T_{\mathrm{c}} \approx 16\,\mathrm{K}$ and is a well established material for superconducting single-photon detectors (SSPDs) \cite{Romestain_2004, Zhang2011, Zhang2017, cheng2020, cucciniello_superconducting_2022}. SSPDs made from NbN and related compounds have been successfully demonstrated on a variety of substrates, including the materials used in this work: LiNbO$_3$ \cite{Tanner_2012, Colangelo:20, sayem2020, Lomonte2021} and GaAs \cite{Gaggero2010, digeronimo2016}. NbN and NbTiN single-photon detectors on LiNbO$_3$-on-insulator waveguides have been demonstrated with system detection efficiencies of 46$\,\%$ and 27$\,\%$, respectively \cite{sayem2020, Lomonte2021}.

Here, we report on a scalable two-stage heterointegration process of III-V epitaxial QD heterostructures and superconducting electrodes on LiNbO$_3$. In this process we adopt the fabrication technologies for NbN SSPDs, epitaxial lift-off, and bonding of III-V heterostructures. The full functionality of the fabricated devices is validated by verifying and quantifying two key performance metrics: (i) the superconducting transition of the electrodes and (ii) the optomechanical and acoustoelectric control of the QDs and simultaneous supercondting operation of the SAW electrodes.\\
The remainder of this article is structured in five sections. Sections 2 and 3 introduce the sample design and experimental techniques employed, respectively. Section 4 validates the electrical performance and superconducting operation. Section 5 reports on three different experimental approaches proving piezo-optomechanical control of QDs in our device. Section 6 summarizes the key findings and gives an outlook on the prospects of our approach and its future applications.

\begin{figure*}[t]
\centering
\includegraphics[width=\textwidth]{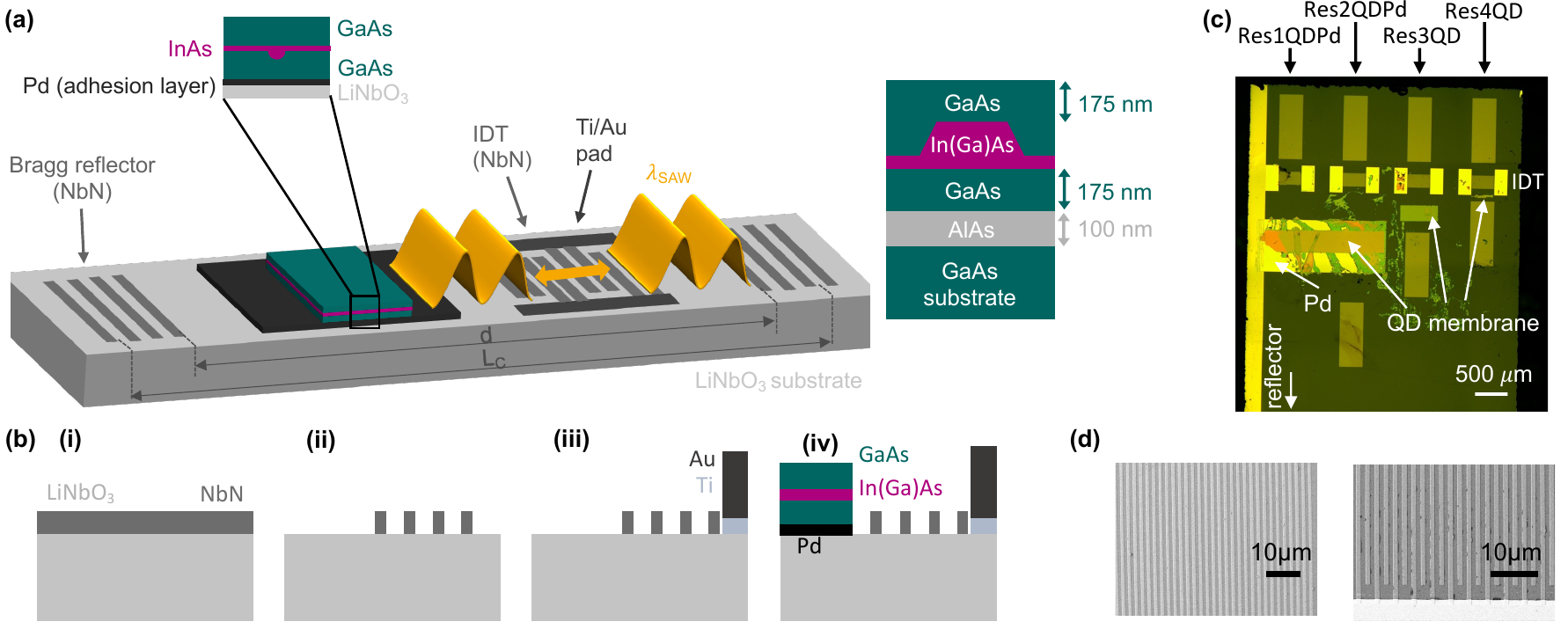}
\caption{\textbf{Device fabrication} -- (a) Schematic of the hybrid device. (b) Heterogeneous integration process flow: (i) NbN sputter deposition, (ii) optical lithography and etch, (iii) contact pad metallization lift-off, (iv) transfer of epitaxial III-V semiconductor heterostructure. (c) Optical microscope image of four devices of different cavity lengths with In(Ga)As QD heterostructure membranes heterointegrated. (d) Scanning electron microscope images of the reflectors (left) and the IDT (right).}
\label{SampleDesignAndFab}
\end{figure*}

\section{Sample design and fabrication}

We implement our two-stage heterointegration on $128^o$ Y-rotated LiNbO$_3$, a common piezoelectric substrate for SAW resonator filters and delay lines with high electromechanical coupling $K^2 =  5.4\,\%$ and phase velocity of $c_0=3978\,\mathrm{m\,s^{-1}}$ at room temperature along the X-direction \cite{morgan_2007}. The prototype device is shown in Figure \ref{SampleDesignAndFab} (a). It is a one-port SAW-resonator aligned along the X-direction with NbN superconducting electrodes and an In(Ga)As QD heterostructure. In essence, two Bragg reflectors separated by a distance $d$ form a SAW resonator with an effective cavity length $L_{\mathrm{c}}$.
An interdigital transducer (IDT) is positioned in close proximity to one of the reflectors for SAW generation. This asymmetric configuration allows for heterointegation of large area semiconductor films. The electrodes of the Bragg reflectors and the IDT are made from NbN. The IDT electrodes are contacted by normal conducting pads. This configuration is adopted from our previous work \cite{Nysten2020} employing normal conducting electrodes which provides the reference for benchmarking the performance of the advanced design developed here. The process flow diagram for sample fabrication is schematically shown in Figure \ref{SampleDesignAndFab}(b). Steps (i) - (iii) are the first heterointegration stage in which the SAW component is realized. It follows a top-down route starting with a continuous thin film and, thus, is fully scalable to the wafer-scale. In step (i) a uniform 20\,nm thick film of NbN is deposited directly on a 100\,mm diameter LiNbO$_3$ substrate using reactive DC magnetron sputtering. The NbN deposition parameters are as follows: 2 mT chamber pressure, 30 sccm argon and 4 sccm nitrogen gas flows, 250 V DC bias voltage. To generate high-quality superconducting NbN films with high $T_c$, an RRR superconducting-grade Nb sputter target was used\cite{astmb393}. In step (ii) this film is subsequentially patterned using optical lithography and inductively coupled plasma reactive ion etching with CF$_{4}$ chemistry. The corresponding ICP etch parameters are the following: 0.4 Pa chamber pressure, 144 sccm CF$_4$ and 9 sccm nitrogen gas flows, 850 W source power, 35 W bias power.  The SAW resonators are finalized in step (iii) when pads composed of Ti (10 nm)/Au (90 nm) are defined by a lift-off process to contact the NbN electrodes of the IDT. The second heterointegration stage is step (iv) during which the semiconductor membrane is transferred onto the LiNbO$_3$ using epitaxial lift-off and transfer \cite{yablonovitch_1987, yablonovitch_van_1990, Nysten2017}.
For the devices we use the heterostructure shown in the right part of Figure \ref{SampleDesignAndFab} (a). It is grown by molecular beam epitaxy on a semi-insulating (001) GaAs substrate starting with a 100\,nm thick AlAs sacrificial layer. The active part of the heterostructure consists of a 250\,nm thick GaAs layer with a single layer of In(Ga)As at its center. This active part is removed from the growth substrate by selectively etching the AlAs sacrificial layer using hydrofluoric acid (HF) and then transferred onto the LiNbO$_3$.

In contrast to our previous work \cite{Nysten2017,Nysten2020}, we study two types of devices. For the first type, shown in Fig. \ref{SampleDesignAndFab} (a), analogous to our previous design, a 50\,nm thick palladium (Pd) adhesion layer on top of a 5\,nm thick titanium (Ti) layer defined via a lift-off process creates a strong and rigid metallurgic bond between LiNbO$_3$ and the semiconductor. This layer also serves to shunt the piezoelectric fields, suppressing acoustoelectric charge carrier dynamics. For the second type, no Ti/Pd is used and consequently the interface between LiNbO$_3$ and the semiconductor is no longer an equipotential plane. Thus, the piezoelectric potential induced by the SAW at the LiNbO$_3$ surface extends into the semiconductor\cite{Weiss2014a}. This allows the verification of piezo-optomechanical coupling comprising dynamic control of the QD energy levels and electric field driven ultrafast carrier dynamics \cite{Schuelein2012, Kinzel2016, Sonner2021} to regulate the charge state of the QD \cite{Schulein2013,Weiss2014b}.
The one-port SAW resonators were designed for a SAW wavelength $\lambda_{0} = 10.0\, \mathrm{\mu m}$ corresponding to a nominal SAW frequency of $f_0\simeq400\,\mathrm{MHz}$. These resonators had four different mirror spacings $d = \{440, 220, 110, 60\}\times\,\lambda_0$ as shown in Fig.\,\ref{SampleDesignAndFab}(c). The effective cavity length is given by $L = d + 2\times L_\mathrm{p}$, where $d$ is the above spacing between the Bragg reflectors and $L_\mathrm{p} = w/r_\mathrm{s} $ is the mirror penetration depth for of a single Bragg reflector. For our electrode width $w = \lambda_{0}/4 = 2.5\,\mathrm{\mu m}$ and single-electrode reflectivity $r_\mathrm{s} \approx 0.02$ \cite{Nysten2020, Manenti2016, morgan_2007}, we obtain $L_\mathrm{p}\approx 130\,\mathrm{\mu m}$. Figure \ref{SampleDesignAndFab} (c) and (d) show optical and scanning electron microscope images of the final devices. Figure \ref{SampleDesignAndFab} (c) shows that Res1QDPd and Res2QDPd contain QD membranes attached via a Pd adhesion layer, whereas the Pd layer is absent for Res3QD and Res4QD.
Table \ref{tab:resparam} gives an overview of the different resonator designs (nominal SAW frequency $f_0$, nominal SAW wavelength $\lambda_0$, mirror spacing $d$, QD membrane, and Pd adhesion layer).
Figure \ref{SampleDesignAndFab} (d) shows high resolution images of the Bragg reflector (left) and the IDT (right) demonstrating successful pattern transfer into the NbN.

\begin{table*}[]
    \centering
    \begin{tabular}{|c|c|c|c|c|c|c|}
        \hline
         Device name & Res1 & Res1QDPd & Res2QDPd & Res2QD & Res3QD & Res4QD \\
         \hline
         $f_0$ (MHz)& 300 & 400 & 400 & 400 & 400 & 400\\
         \hline
         $\lambda_0$ (µm) & 13.3 & 10.0 & 10.0 & 10.0 & 10.0 & 10.0\\
         \hline
         $d$ $(\lambda_{0})$ & $340$ & $440$ & $220$ & $220$& $110$ & $60$ \\
         \hline
         $d$ (µm) & 4522 & 4400 & 2200 & 2200 & 1100 & 600\\
          \hline
         QD membrane & no & yes & yes & yes & yes & yes\\
         \hline
         Pd layer & no &yes & yes & no & no & no \\
         \hline
    \end{tabular}
    \caption{Overview of the different resonator designs, where $f_0$ nominal SAW frequency, $\lambda_0$ is the SAW wavelength as defined lithographically, $d$ is the mirror spacing defined lithographically, and the bottom two rows indicate the presence or absence of a QD membrane and a Pd adhesion layer.}
    \label{tab:resparam}
\end{table*}

\section{Characterization techniques}

The sample was mounted on a custom-made carrier and wire-bonded. All experiments were conducted with the sample loaded into a variable temperature closed-cycle optical cryostat equipped with radio frequency (rf) signal lines. For rf characterization a vector network analyzer (VNA) was used to determine the $S_{11}$ scattering parameter at different applied rf power levels $(P_{\mathrm{rf}})$.
The piezo-optomechanical response of single QDs was characterized by a combination of time-integrated and time-resolved photoluminescence (PL) spectroscopy \cite{Weiss2018}. Here, electrical excitation of the SAW resonator was performed using an rf signal generator. A directional bridge was used to analyze the reflected electrical power using an oscilloscope online during the optical experiment to ensure consistency between the optical and rf electrical data \cite{Nysten2020}.
In the experiments presented here, optical excitation was performed using a continuous wave $\lambda=833\,\mathrm{nm}$ diode laser focused to a diffraction limited spot with a diameter of $<1.5\,\mathrm{\mu m}$ by a $NA=0.81$ microscope objective with a focal length of 2.89\,mm. A moderate optical pumping density $P_{\mathrm{optical}}\approx 191 \,\mathrm{W}/\mathrm{cm^2}$ ensured preferential generation of single exciton species with no noticeable signatures of biexcitons. The emission of the QDs was collected by the same objective and dispersed in a 0.7\,m grating monochromator. Time-integrated multi-channel detection was performed using a cooled CCD detector. For single-channel time-resolved detection, the detection wavelength was scanned and the signal was recorded by a single photon avalanche diode (SPAD) with time resolution $< 50$\,ps. The electrical pulses of the SPAD were recorded with time correlation electronics referenced to the electrical signal applied to the IDT \cite{Buehler2022,Kinzel2011}. 

\label{sec:sample}
\section{Electrical characterization}

\begin{figure*}[t]
\centering
\includegraphics[width=\textwidth]{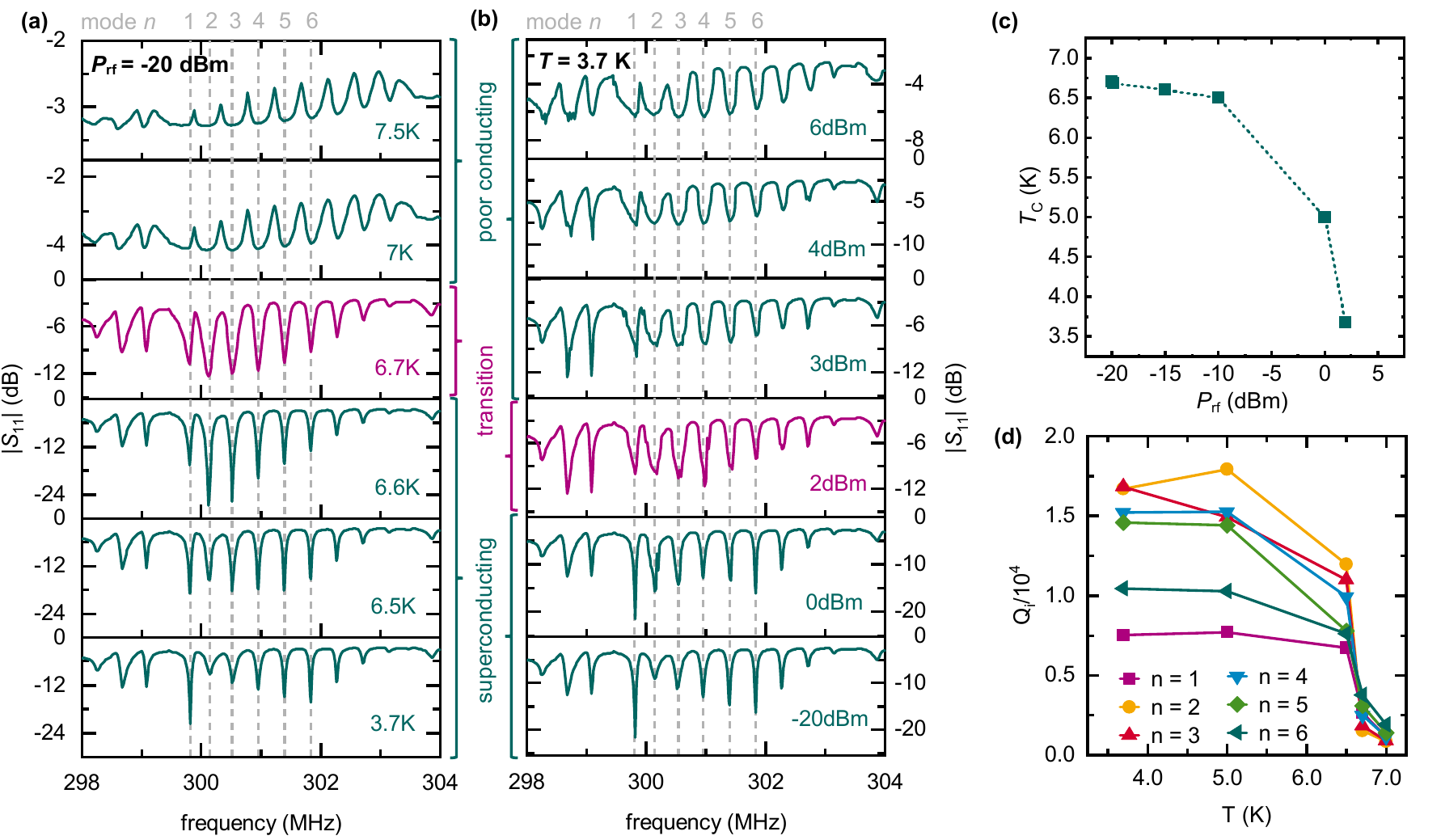}
\caption{\textbf{Electrical characterization} -- (a) Scattering parameter $S_{11}$ of a $f_0=300\,\mathrm{MHz}$ SAW resonator at $P_{\mathrm{rf}}=-20\,\mathrm{dBm}$ with temperature $T$ increasing from bottom to top and (b) at $T=3.7$ $\mathrm{K}$ with $P_{\mathrm{rf}}$ increasing from bottom to top. (c) Phase diagram in the $P_{\mathrm{rf}}$-$T$ parameter space indicating the phase boundary from the superconducting to normal conducting state.
(d) Internal quality factor $Q_i$ of the phononic modes as a function of $T$.}
\label{fig:Fig2}
\end{figure*}

We first validate the electrical operation of our device in the superconducting regime of NbN and determine the accessible ranges of the main operation parameters, the sample temperature, $T$, and the rf electrical power, $P_{\mathrm{rf}}$, applied to the IDT. Figure \ref{fig:Fig2} shows a comprehensive set of data and their analysis for resonator Res1 before transfer of the QD heterostructure. The nominal design frequency was $f_0=300\,\mathrm{MHz}$ with a nominal mirror distance of $d_1 = 340\lambda_0$ with $ \lambda_0 = 4522\,\mu\mathrm{m}$.  
We begin with applying a constant low rf power $P_{\mathrm{rf}}=-20\,\mathrm{dBm}$ and measure the $S_{11}$ scattering parameter as a function of the applied electrical frequency for different values of the sample temperature in the cryostat. These data are plotted in Figure \ref{fig:Fig2} (a) with $T$ increasing from the bottom to top spectrum. The spectrum recorded at the base temperature $T=3.7\,\mathrm{K}$ shows the expected resonator response with clearly resolved minima in the reflected rf power at the resonance frequencies of the cavity modes (vertical dashed lines). The measured free spectral range $\mathrm{FSR}=(410\pm 40)\,\mathrm{kHz}$ corresponds to an effective resonator length $L_{\mathrm{c}}=(4866\pm 568)\,\mathrm{\mu m}$, in good agreement with the nominal lithographically defined cavity length $d = 4522\,\mathrm{\mu m}$. Second, the measured $S_{11}$ can be fitted using
\begin{equation}
	\label{Equ:S11}
	S_{11}(f)=\frac{(Q_{\mathrm{e},n}-Q_{\mathrm{i},n})/Q_{\mathrm{e},n}+2iQ_{\mathrm{i},n}(f-f_n)/f}{(Q_{\mathrm{e},n}+Q_{\mathrm{i},n})/Q_{\mathrm{e},n}+2iQ_{\mathrm{i},n}(f-f_n)/f}
\end{equation}
with $Q_{\mathrm{i},n}$ and $Q_{\mathrm{e},n}$ being the internal and external Q-factors of mode $n$ at frequency $f_n$ \cite{Manenti2016}.
At base temperature, we obtain values ranging between $7000\leq Q_{\mathrm{i},n}\leq 17000$, indicating a three-fold improvement compared to identical resonator devices equipped with normal conducting Ti/Al electrodes \cite{Nysten2020}. 

As $T$ increases, we observe the transition from superconductivity to normal conductivity at $T_c=6.7\,\mathrm{K}$ (purple spectrum). For temperatures above $T_c$, we observe a characteristic change of the spectrum with the emergence of peaks (instead of dips) that are shifted in frequency compared to the modes in the superconducting state (dashed lines). This behavior is expected for loss-dominated resonators, which is the case for NbN thin films above $T_c$. Its conductivity is at least one order of magnitude lower than commonly used normal conducting metals like aluminum \cite{Gavaler1971, wang1996, morgan_2007}.
In Figure \ref{fig:Fig2} (b) we show analogous $S_{11}$ spectra with the temperature held constant at $T=3.7\,\mathrm{K}$ and the electrical power applied to the IDT increasing from the bottom to the top. As $P_{\mathrm{rf}}$ increases from $P_{\mathrm{rf}}=-20\,\mathrm{dBm}$, we observe a clear change of the spectrum at $P_{\mathrm{rf}}=+2\,\mathrm{dBm}$ (purple). This change is markedly different to the abrupt change in Figure \ref{fig:Fig2} (a) between $T_c=6.7\,\mathrm{K}$ and $T=7\,\mathrm{K}$. Here, the dips at the resonator mode frequencies broaden continuously and evolve in a spectrum similar to the normal conducting state at the highest power level $P_{\mathrm{rf}}=+6\,\mathrm{dBm}$. This continuous broadening cannot be explained by a change of the global sample temperature, but instead indicates a local breakdown of superconductivity in a subset of the electrodes. Since the cavity mode frequencies (vertical grey lines) remain constant, we conclude that superconductivity is initially preserved in the Bragg mirrors' electrodes. Breakdown occurs locally in the IDT electrodes, where the driving rf electrical signals drives a current. The corresponding local increase of the current density above the critical value breaks superconductivity. This occurs first at the cavity resonance where the highest current densities are reached. With increasing $P_{\mathrm{rf}}$, the critical current density is reached off-resonance giving rise to the observed apparent broadening. At the highest $P_{\mathrm{rf}}$, superconductivity breaks down over the entire range of frequencies. Ohmic heating in this normal conducting state raises the global sample temperature above $T_c$, retaining a spectrum similar to that observed in Figure \ref{fig:Fig2} (a).

To further corroborate this interpretation we analyze the observed $T_c$ as a function of $P_{\mathrm{rf}}$, plotted in Figure \ref{fig:Fig2} (a). These data show that an increase of $P_{\mathrm{rf}}$ leads to a decrease of the measured $T_c$. At this point we bear in mind that an increase of $P_{\mathrm{rf}}$ corresponds to an increase of the current density. Thus the data in \ref{fig:Fig2} (a) corresponds to an effective phase boundary of a superconductor in the current-temperature parameter space. Finally, we analyze $Q_{\mathrm{i},n}$ as a function of $T$ obtained from a best fit of Equation \ref{Equ:S11} to the data in Figure \ref{fig:Fig2} (a). The obtained values for modes $n=1,2,\dots6$ are plotted for temperatures $3.7 \leq T\leq 7\,\mathrm{K}$ in Figure \ref{fig:Fig2} (d). The data clearly show that high $Q_{\mathrm{i},n}$ is in fact preserved for all modes up to $T_c$ when superconductivity breaks down. To summarize, the performed electrical characterization proves superconducting operation  with high internal $Q_{\mathrm{i},n}> 17000$. The derived effective phase boundary defines the parameter space for the operation of our device. For all experiments shown in the remainder of this article, the effective phase boundary was first pinpointed to ensure superconducting operation in the optical experiments.

\section{Piezo-optomechanical characterization}

In the second phase of our characterization, we investigate the piezo-optomechanical interaction between the SAW resonator and the QDs in the heterointegrated device. We probe the optomechanical coupling between single QDs and phononic modes by measuring the optical emission as a function of the applied rf parameters. In all these experiments, the electrodes are in the superconducting state. Thus we validate \emph{combined} superconducting operation and SAW control of single QDs. In the following, the design frequency of the investigated devices is $f_0 \approx 400\,\mathrm{MHz}$. In Sections \ref{subsec:timeintegratedPL} and \ref{subsec:timeresPL}, we use the resonator Res2QDPd resonator with the QD heterostructure transferred onto a Pd adhesion layer in its center. The Pd shunts the electric field induced by the SAW \cite{Pustiowski2015, Nysten2017}. Accordingly, Stark-effect modulation, which becomes dominant at high SAW amplitudes \cite{Weiss2014a} and acoustoelectric charge state regulation \cite{Volk2010,Schulein2013} are effectively suppressed and can be neglected in the following. In Section \ref{subsec:AEEPL} we use a Res2QD resonator where a QD heterostructure directly transferred onto the LiNbO$_3$ without a Pd adhesion layer. In this case, the piezoelectic fields can induce spatio-temporal carrier dynamics and thus, regulate the charge state of the QD.

\subsection{Time-integrated optomechanical characterization}
\label{subsec:timeintegratedPL}

\begin{figure*}[t]
\centering
\includegraphics[width=\textwidth]{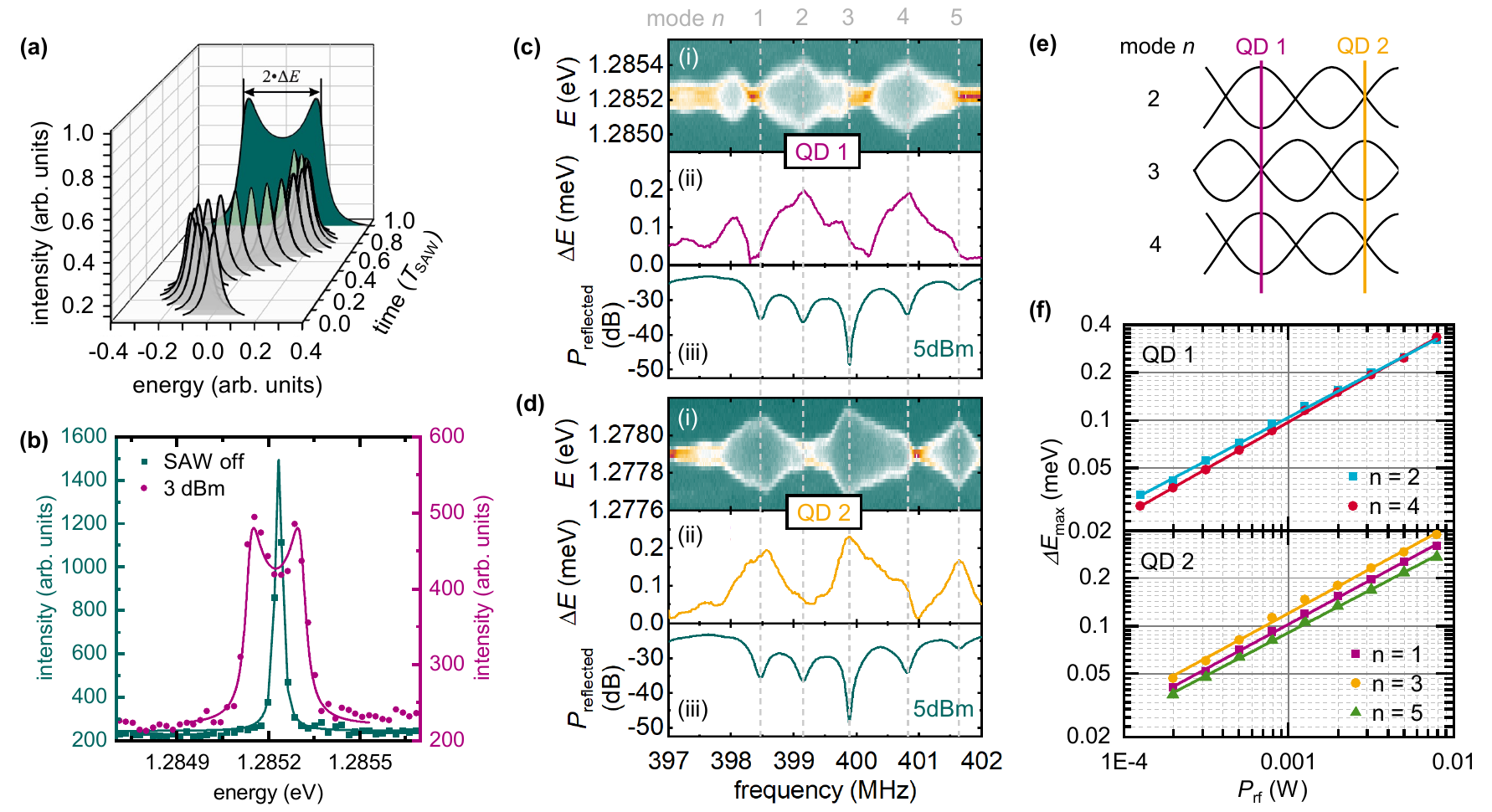}
\caption{
\textbf{Piezo-optomechanical characterization} -- (a) Schematic of dynamic modulation of Lorentzian emission line giving rise to the time-integrated spectrum given by \ref{Equ:LorentzSine}. (b) Emission line of a single QD (data points) without a SAW applied (green) and with a SAW applied (purple) with best fits of Equation\,\,\ref{Equ:LorentzSine} (solid lines). (c) and (d) PL spectrum (\textit{i}), extracted $\Delta E$ (\textit{ii}) and reflected electrical rf power (\textit{iii}) recorded from QD\,1 and QD\,2, respectively. (e) Schematic of site-selective coupling of QD\,1 and QD\,2. (f) $\Delta E$ for different phononic modes as function of $P_{\mathrm{rf}}$ of QD\,1 (top) and QD\,2 (bottom).
}
\label{fig:Fig3}
\end{figure*}

We first measure the optomechanical response of a single QD by time and phase averaged PL spectroscopy. As schematically shown in Figure \ref{fig:Fig3} (a), the detected line shape is a time-average of the sinusoidally modulated Lorentzian QD emission line \cite{Weiss2014a, Weiss2017} given by

\begin{equation}
\label{Equ:LorentzSine}
	I (E)=I_0 + f_{\mathrm{rf}}\cdot \frac{2A}{\pi}\cdot
    \int_{0}^{\frac{1}{f_{\mathrm{rf}}}}\frac{\omega}{4( E-  (   E_0 + \Delta E \cdot \sin(2\pi f_\mathrm{rf}\cdot t) ) )^2 + \omega^2}\mathrm{d}t \ .
\end{equation}

In this expression, $I_0$ denotes a time-independent intensity offset, $E_0$ the center energy, $A$ the amplitude, $\omega$ the width of the Lorentzian emission peak, and $\Delta E$ the optomechanical modulation amplitude.
Figure\,\,\ref{fig:Fig3}(a) schematically shows the sinusoidal modulation of the Lorentzian emission line of a single QD (gray) and the resulting time-integrated emission spectrum (green) with its optomechanical tuning amplitude labeled as $2\Delta E$.
Figure\,\,\ref{fig:Fig3}(b) shows the measured time-integrated emission spectra of a exemplary QD (data points) and a best fit of Equation\,\,\ref{Equ:LorentzSine} (solid line). Without a SAW applied (green), the expected Lorentzian line is faithfully detected. When strained by a SAW ($P_{\mathrm{rf}} = 3\,$dBm, $f_{\mathrm{rf}} = 399.62\,$MHz, purple), the characteristic lineshape given by Equation\,\,\ref{Equ:LorentzSine} is observed. 

Next, we confirm the optomechanical coupling of QDs to the SAW resonator modes. To this end, we begin by applying $P_{\mathrm{rf}}=5\,$dBm to the IDT, at which the resonator was verified to be in the superconducting state. We then scan $f_{\mathrm{rf}}$ from 397 to 402\,MHz. We select two QDs, QD\,1 and QD\,2, which are separated by $\approx 1.5\times\lambda_0$ along the axis of the resonator.
The results of the performed characterization experiments are shown for QD\,1 and QD\,2 in Figure\,\,\ref{fig:Fig3}(c) and (d), respectively.
The upper panels (\textit{i}) show the time-integrated PL intensity as a function of electrical $f_{\mathrm{rf}}$ (horizontal axes) applied to the IDT and photon energy (vertical axes). The lower panels (\textit{iii}) show the simultaneously recorded reflected rf power to identify the involved SAW modes $f_{n}$ of mode index ($n$) of the hybrid SAW resonator. These are labelled and marked by the vertical dashed lines.
The center panels (\textit{ii}) show $\Delta E (f_{\mathrm{rf}})$ of the two QDs extracted from the experimental data by fitting Equation\,\,\ref{Equ:LorentzSine}.
These data clearly prove the anticipated site and frequency selective coupling of the embedded QDs and the phononic modes. QD\,1 in Figure\,\,\ref{fig:Fig3} (c) shows strong optomechanical response $\Delta E$ and, thus, strong optomechanical coupling when $f_{\mathrm{rf}}$ is in resonance with an even index mode $n = 2, 4$. Conversely, this coupling is suppressed for odd index models $n = 1, 3, 5$. The optomechanical coupling is inverted for QD\,2 in Figure\,\,\ref{fig:Fig3} (d), which exhibits strong and weak coupling for $n = 1, 3, 5$  and $n = 2, 4$, respectively. These observations are in agreement with the QD\,1 being at the antinodes of the $n = 2, 4$ modes and QD\,2 being at those of the $n = 1, 3, 5$ modes as shown schematically in Figure\,\,\ref{fig:Fig3} (e) \cite{Nysten2020}.

In a third step, we study the dependence of the optomechanical modulation $\Delta E$ on the applied rf power. We increase $P_{\mathrm{rf}}$ from $P_{\mathrm{rf}}=-9\,\mathrm{dBm}$ in steps of $2\,\mathrm{dB}$ to $+9\,\mathrm{dBm}$ and record the resulting $\Delta E$ of QD\,1 and QD\,2. In this range of $P_{\mathrm{rf}}$, the device is in the superconducting state, which was confirmed by the simultaneously measured electrical power reflected from the device. This measured range is $3\,\mathrm{dB}$ larger than that of the $300\,\mathrm{MHz}$ device shown in \ref{fig:Fig2}.
The extracted $\Delta E$ of QD\,1 and QD\,2 are plotted as a function of $P_{\mathrm{rf}}$ in the upper and lower panel of Figure\,\,\ref{fig:Fig3}(f), respectively. These data are presented for all modes to which the respective QDs couple. Moreover, the data is plotted in double-logarithmic representation to identify the power law dependence $\Delta E \propto P_{\mathrm{rf}}^m$ and the underlying coupling mechanism. 
QD\,1 exhibits a power law dependence with an average slope $m_{\mathrm{QD1}}=0.57\pm 0.01$ and QD\,2 with $m_{\mathrm{QD2}}=0.55\pm 0.01$. The amplitude of the SAW, $u_z\propto \sqrt{P_{\mathrm{rf}}}$, and $\Delta E\propto u_z$. Hence, a slope of $m\sim0.5$ is a characteristic fingerprint for deformation potential coupling being the dominant mechanism \cite{Weiss2018}. We note that the small increase compared to the ideal value of $m=0.5$ may arise from a weak nonlinearity as observed for similar hybrid devices with normal conducting electrodes \cite{Nysten2020}. Moreover, the extracted slope excludes Stark effect modulation for which $m=1$ is expected \cite{Weiss2014a}.

\subsection{Time-resolved optomechanical characterization}
\label{subsec:timeresPL}

\begin{figure}[t]
    \centering
    \includegraphics[width=0.5\textwidth]{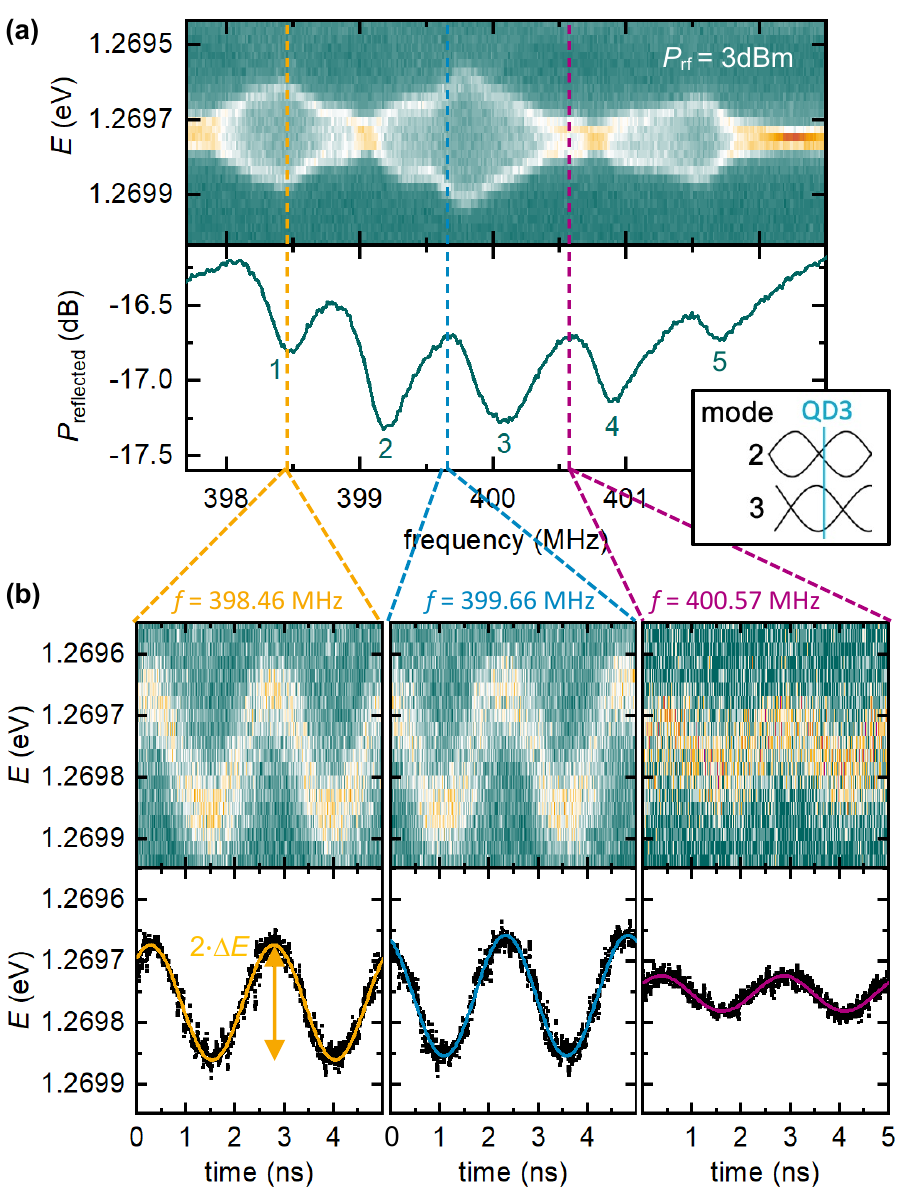}
    \caption{
    \textbf{Time-resolved dynamic modulation} -- (a) Time integrated PL spectrum of QD\,3 and reflected rf power. The inset shows the position of QD with respect to the $n=2$ and $n=3$ modes. (b) Time-dependent PL spectra (top) for three selected $f_{\mathrm{rf}}$ marked in (a) and extracted spectral modulations (bottom). 
    }
    \label{fig:Fig4}  
\end{figure}

Next, we perform time-correlated single-photon counting to resolve the SAW-driven dynamics of QD\,3 directly in the time domain. This type of characterization allows direct observation of the temporal shift of the QD emission line, which was not addressed in our previous work on hybrid QD-SAW resontors with normal conducting electrodes \cite{Nysten2020}. In the following, $P_{\mathrm{rf}}=3\,$dBm is fixed. Before conducting these time-resolved measurements, we study the time-integrated emission of QD\,3 as in the previous section to determine the position of the selected QD in the SAW cavity field.
Figure\,\,\ref{fig:Fig4}(a) shows the time-integrated emission spectrum of QD\,3 (top panel) and the simultaneously measured reflected rf power (bottom panel) as a function of $f_{\mathrm{rf}}$.
From these data we conclude that QD\,3 couples strongly to odd modes and only weakly to even modes.
The derived relative position of QD\,3 within the SAW cavity field is shown schematically as an inset of Figure\,\,\ref{fig:Fig4}(a).
In addition and in agreement with the data shown in Figure\,\,\ref{fig:Fig3} and \cite{Nysten2020}, a pronounced optical response can be found in the frequency range between the $n=2$ and $n=3$ modes. 
The time-resolved analysis of the dynamically strained QD\,3 is performed for three characteristic frequencies $f_{\mathrm{rf}}$ marked by different colored lines in Figure\,\,\ref{fig:Fig4} (a).
$f_1=398.46\,$MHz corresponds to the resonance of mode $n=1$. $f_{23}=399.66\,$MHz and $f_{34}=400.57\,$MHz are chosen in between the $n=2$/$n=3$ and $n=3$/$n=4$ modes, respectively. At these frequencies, QD\,3 shows strong ($f_{23}$) and weak ($f_{34}$) modulations.
The top panels of Figure\,\,\ref{fig:Fig4} (b) present plots of the temporal modulation of the QD emission line at the three selected frequencies in false color representation. In all three cases, a clear sinusoidal modulation is observed in the time domain. Furthermore, the period of these modulations correspond to that set by the applied rf.
The bottom panels of Figure\,\,\ref{fig:Fig4} (b) show the peak positions (black) as a function of time extracted from these data by best fits of a Lorentzian line for each time stamp.
By fitting the obtained data with a sine function (solid orange, blue, and purple lines) we extract modulation frequencies of the QD $f_{\mathrm{QD,1}}=398.45\pm2.77\,\mathrm{MHz}$, $f_{\mathrm{QD,23}}=399.66\pm2.75\,\mathrm{MHz}$, and $f_{\mathrm{QD,34}}=400.45\pm 1.27\,\mathrm{MHz}$. These values agree well with the electrical rf. Thus, these data prove that in this sample the QD is predominantly modulated at the SAW excitation frequency, regardless of its location in the phononic mode spectrum. These findings furthermore clearly indicate that no dominant wave mixing processes occur in the SAW resonator and that the dominant coupling mechanism is deformation potential coupling.

\subsection{Acoustoelectric charge state control}
\label{subsec:AEEPL}
\begin{figure}[htbp]
\centering
\includegraphics[width=0.5\textwidth]{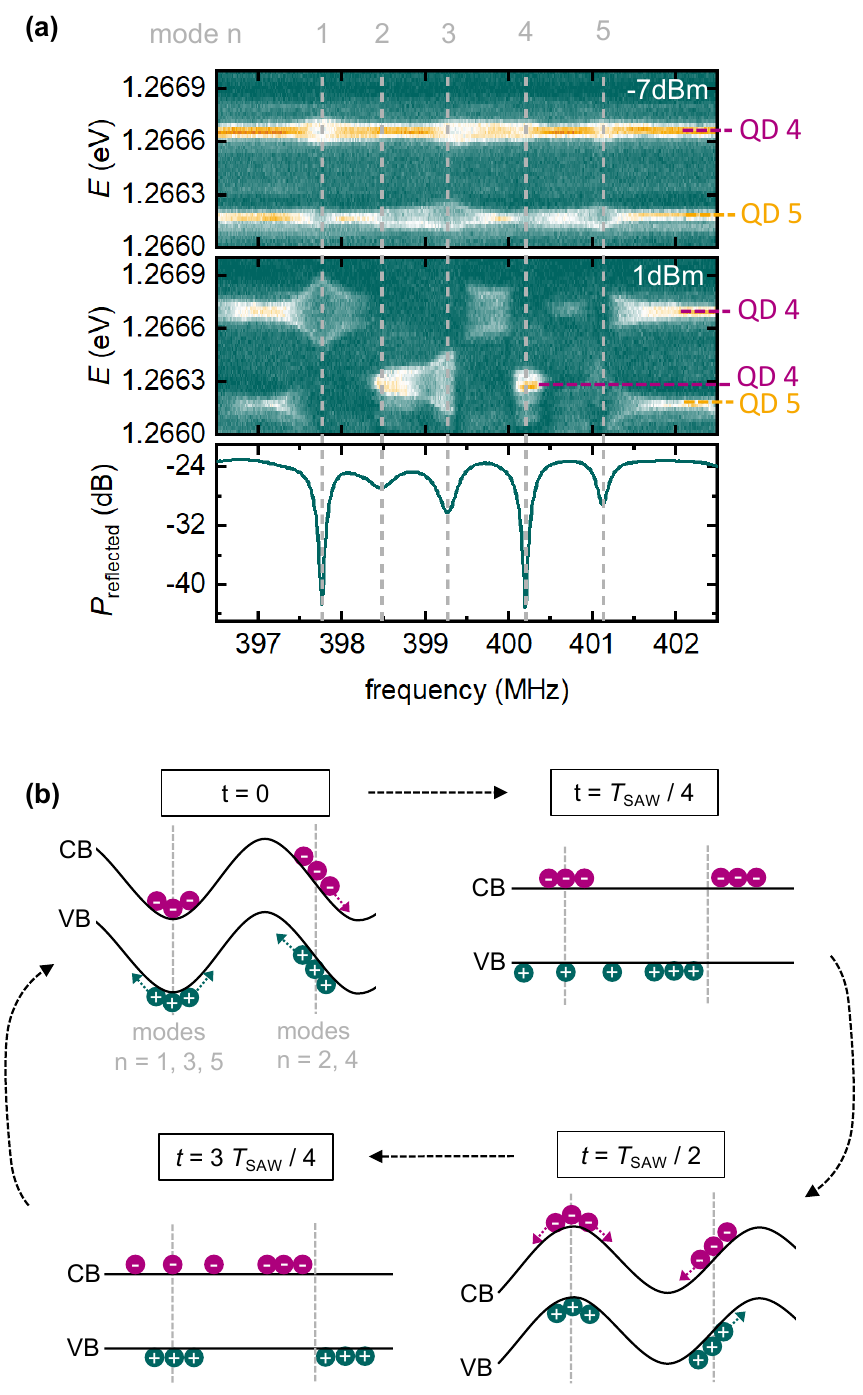}
\caption{
\textbf{Acoustoelectric charge state regulation} -- (a) PL spectra as function of $f_{\mathrm{rf}}$ at $P_{\mathrm{rf}} = -7\,\mathrm{dBm}$ (top) and $P_{\mathrm{rf}} = +1\,\mathrm{dBm}$ (center) showing the emission of QD\,4 and QD\,5 and simultaneously measured reflected rf power (bottom). (b) Schematic SAW-induced bandstructure modulation in the device and the resulting acoustoelectric electron (purple) and hole (green) dynamics for one full acoustic cycle. Dashed lines mark the position of QD\,4 when even and odd index modes are excited. 
}
\label{fig:Fig5}
\end{figure}

Finally, we demonstrate acoustoelectric control of the charge states of single QDs on our hybrid platform. To this end, we study resonator Res2QD with a nominal SAW frequency of $f_0=400\,\mathrm{MHz}$ with a mirror spacing d = 220 $\lambda_0$ and a QD heterostructure directly transferred onto the LiNbO$_3$ without a Pd adhesion layer. Again, the design frequency is $f_0=400\,\mathrm{MHz}$ as confirmed by the SAW mode spectrum plotted in the lower panel of Figure\,\,\ref{fig:Fig5} (a). The upper panel of Figure\,\,\ref{fig:Fig5} (a) shows the emission spectrum of two QDs, QD\,4 and QD\,5 (plotted in false-color representation as a function $f_{\mathrm{rf}}$). These data are recorded at relatively low rf power of $P_{\mathrm{rf}} = -7\,\mathrm{dBm}$ at which strain coupling is dominant \cite{Weiss2014a,Weiss2014b}. Thus, both emission lines QD\,4 at $E=1266.66\,\mathrm{meV}$ and QD\,5 at $E_5=1266.17\,\mathrm{meV}$ show the expected broadening when $f_{\mathrm{rf}}$ is tuned into resonance with modes of odd ($n=1,3,5$) index, while no significant coupling is observed for modes of even index ($n=2,4$). Thus, we can conclude that both QDs are at the antinodes (nodes) of the strain field of odd (even) index modes. This observation proves mode-selective \emph{strain} coupling of the QD to the SAW even with a relatively weak van der Waals bond between the semiconductor and LiNbO$_3$ substrate, compared to the rigid metalurgical bond for the devices with the Pd adhesion layer.
Next, we increase the rf power to $P_{\mathrm{rf}} = +1\,\mathrm{dBm}$ and plot the recorded emission spectra in the same range of photon energies and $f_{\mathrm{rf}}$ in false-color representation in the center panel of Figure\,\,\ref{fig:Fig5} (a). These data exhibit a completely different behaviour than those of samples with a Pd adhesion layer [cf. Figures \ref{fig:Fig3} and \ref{fig:Fig4}, and references \cite{Nysten2017,Nysten2020}] arising from the combination of strain tuning and acoustoelectrically driven carrier dynamics by the SAW. In this device without the Pd adhesion layer, no highly-conductive metal shortens the piezoelectric fields induced by the SAW on the LiNbO$_3$-GaAs interface. Thus, the electric field extends into the semiconductor. This field efficiently ionizes the photogenerated excitons and induces spatio-temporal charge carrier dynamics (STCDs) \cite{Garcia2014,Schuelein2012,Kinzel2016}. The induced dynamics regulate the charge state of the QDs on timescales of the SAW and lead to correlated suppression and emergence of different emission lines \cite{Volk2010,Schulein2013}. We note that in contrast to other previous work, these dynamics are observed within a SAW resonator, and no freely propagating SAWs were employed.

In the following analysis, we focus on QD\,4 for which different emission lines can be clearly identified. In order to understand the experimental findings, we have to consider that our device is a SAW resonator in contrast to devices with propagating SAWs studied previously in literature. In our resonator, the nodes of the phononic modes' standing wave pattern are stationary. To understand the experimental findings, we have to consider the time-dependent strain and electric field at the position of QD\,4 for a given mode index $n$. As shown above, QD\,4 is located at the antinodes of odd index ($n=1,3,5$) modes and at nodes of of even index ($n=2,4$) modes. For X-propagating SAWs on $128^o$ Y-rotated LiNbO$_3$, the volume dilatation inducing the optomechanical modulation and the electric potential are in phase and simultaneously tune the emission line and regulate the occupancy state. Figure\,\,\ref{fig:Fig5} (b) schematically depicts the dynamic evolution of the bandstructure modulation at four distinct times during the acoustic cycle. At $t=0$, the amplitude of piezoelectric potential of the SAW is maximum. Thus, the corresponding sinusoidal modulation is superimposed giving rise to the well established type-II band-edge modulation \cite{Rocke1997}. At $t=T_{\mathrm{SAW}}/2$, the situation is reversed and positions of maxima and minima are exchanged. At $t=T_{\mathrm{SAW}}/4$ and $t=3T_{\mathrm{SAW}}/4$, destructive interference of the SAW fields occurs and leads to an unperturbed flat bandstructure. In these schematics, the vertical dashed lines indicate the position of QD\,4 in this bandstructure when odd index ($n=1,3,5$) modes or even index ($n=2,4$) modes are excited.

For $n=2,4$, QD\,4 is at the node of the electric potential modulation. Thus, the resulting gradient and hence amplitude of its electric field is maximum. This leads to an efficient dissociation of excitons and pronounced STCDs as shown in the schematics of Figure\,\,\ref{fig:Fig5} (b). In the experimental data, we observe a strong suppression of the emission line observed for the weakly modulated case at $E=1266.66\,\mathrm{meV}$ and a new emission line at $E=1266.28\,\mathrm{meV}$. This switching is a characteristic fingerprint of acoustically regulated carrier injection driven by STCDs. For $n=1,3,5$, QD\,4 is at the antinode of the electric potential modulation. Thus, the gradient and electric field vanishes and the STCDs are dominated by redistribution processes of electrons and holes from their unstable points (and position of the QD) to regions of maximum electric field. These processes are indicated in the schematics of Fig.\,\ref{fig:Fig5} (b) and are slow compared to field-driven drift. This leads to marked changes in the carrier injection dynamics into QD\,4 which favors the preferential generation of different occupancy states and resulting emission lines for different $n$. For $n=1$, the occupancy state corresponding to the $E=1266.66\,\mathrm{meV}$ line is preferentially generated, while for $n=3$, that of the $E=1266.28\,\mathrm{meV}$ line. For $n=5$, both lines are almost completely suppressed which points towards efficient carrier depletion at the position of the QD\,4. When tuning $f_{\mathrm{rf}}$, a characteristic and reproducible switching pattern is observed. This observation unambiguously proves that the direct coupling of QD heterostructure to the LiNbO$_3$ substrate leads to pronounced STCDs and charge state regulation which can be efficiently suppressed by a thin metallic layer shunting the electric fields. Note that detailed modelling at the level possible for propagating SAWs \cite{Schulein2013,Weiss2014b} is not possible for the devices studied here. As shown for the piezo-optomechanical response the complex mode pattern of our resonator comprises contributions of propagating and stationary waves. The observed switching hampers the faithful disentanglement of these contributions, which would be required to perform numerical simulations of the STCDs.

\section{Conclusion}
In summary, we developed and implemented a two-step heterointegration process of a hybrid SAW-resonator device comprising superconducting electrodes and an epitaxial semiconductor heterostructure on a LiNbO$_3$ substrate. Our facile process can be scaled to the wafer-scale by building on recent breakthroughs in this field. Firstly, the here applied transfer of millimeter-sized and few 100 nm-thick semiconductor heterostructure membranes can be realized at the wafer-scale through wafer bonding techniques \cite{Hackett2021, Hackett2023, Cheng2013}. Secondly, superconducting thin films are sputter-deposited and patterned in a subtractive process, which is in principle also directly scalable. We validated the functionality of the fabricated devices by characterizing the parameter space for superconducting operation of the SAW component. In these first experiments, we monitored the superconducting to normal conducting transition of the NbN electrodes as a function of temperature and applied electrical rf power. The achieved internal quality factor $Q_i\approx17000$ marks a three-fold improvement to previously studied similar devices using normal conducting electrodes \cite{Nysten2020}. After transfer of the III-V semiconductor, we verified the combined superconducting operation of the SAW device and piezo-optomechancial control of the embedded QDs. In a series of experiments, mode-selective coupling of the QDs, time-modulation of the QD emission line, and acoustically regulated carrier injection are verified. The latter observation provides direct evidence of spectral tuning of the QD by dynamic strain and simultaneous acoustoelectric regulation of the QD's occupancy state for the first time in a SAW resonator.
\\

Our two-step heterointegration opens directions for advanced piezo-optomechanical quantum devices. First, our superconducting SAW resonators are fabricated with the same processes used for NbN single-photon detectors \cite{Lomonte2021}, enabling future SAW devices to be integrated with superconducting single-photon detectors during a single fabrication step. Second, the performance of our resonators can be deliberately enhanced to small mode volume ($V \approx \lambda^3$),  higher Q-factor, and high frequency ($>$ 1 GHz) operation \cite{Shao2019a}. These devices then harness the large $K^2$ of LiNbO$_3$ and strong optomechanical coupling of III-V semiconductor QDs \cite{Metcalfe2010, Weiss2021, Wigger2021a}. These may ultimately enable coherent optomechanical control in the sideband regime \cite{imany2022quantum,hahn2022photon}. Third, the process can be extended to additional heterointegration steps for example adding defect quantum emitters providing spin qubits serving as quantum memories \cite{maity_coherent_2020,Golter2016a,Sukachev2017,Neuman2021}.

\newpage

\section*{Acknowledgements}
We gratefully acknowledge support via the UC Santa Barbara NSF Quantum Foundry funded via the Q-AMASE-i program under award DMR-1906325. This project was funded by the Deutsche Forschungsgemeinschaft (DFG, German Research Foundation) - 465136867 and the Bavaria-California Technology Center BaCaTeC (F\"{o}rderprojekt Nr. 8 [2020-1]). K.M. acknowledges financial support via the German Federal Ministry of Education and Research (BMBF) via the funding program Photonics Research Germany (contract number 13N14846). M.L. and J.J.F. acknowledge funding by the Bavarian Hightech Agenda within the Munich Quantum Valley doctoral fellowship program. J.J.F. and K.M. acknowledge financial support by the Deutsche Forschungsgemeinschaft (DFG, German Research Foundation) under Germany’s Excellence Strategy—EXC-2111—390814868. We thank Hubert Riedl (WSI-TUM) for crystal growth.\\

\bibliographystyle{iopart-num}
\bibliography{bibliography_clean}

\end{document}